\documentclass[review]{article} \usepackage{natbib}
\usepackage{graphicx} 
\usepackage{amsfonts,amsmath,amssymb}
 \usepackage{subfigure}
\usepackage{hyperref}
\pdfoutput=1
\begin{document}
\title{Comparisons of Hyv\"arinen and pairwise
  estimators  in two simple linear time series models} 
\author{{\large V.~Mameli\thanks{mameli.valentina@virgilio.it} (1),  M.~Musio\thanks{mmusio@unica.it} (1), A.~P.~Dawid\thanks{ apd@statslab.cam.ac.uk} (2)}\\
       {\small ((1) Department of Mathematics and Computer Science, University of Cagliari, Italy,}\\ {\small (2) Centre for Mathematical Sciences, University of Cambridge, UK)}}
       \date{}
\newcommand{\Real}{{\rm I}\negthinspace {\rm R}}

\maketitle
\begin{abstract}
  The aim of this paper is to compare numerically the performance of
  two estimators based on Hyv\"arinen's local homogeneous scoring rule
  with that of the full and the pairwise maximum likelihood
  estimators.  In particular, two different model settings, for which
  both full and pairwise maximum likelihood estimators can be
  obtained, have been considered: the first order autoregressive model
  (AR(1)) and the moving average model (MA(1)).  Simulation studies
  highlight very different behaviours for the Hyv\"arinen scoring rule
  estimators relative to the pairwise likelihood estimators in these
  two settings. 
  \\

\end{abstract}

\section{Introduction}
\label{intro}
Recent years have seen growing interest in composite likelihood
methods, due to their computational advantages in estimating
parameters of very complex statistical models: see \citet{Varin:2011}
for an overview.  A key feature of these methods is their ability to
avoid the calculation of the normalizing constant of the model, which
will typically depend on the parameter.  Determination of this
constant, essential for full likelihood-based inference, can be a very
challenging task, entailing multidimensional integration of the full
joint density.  Composite likelihood approaches can avoid this, by
maximizing the product of low-dimensional marginal or conditional
likelihoods.  The most used composite likelihood in applications is
the pairwise likelihood \citep{Cessie}, defined as the product of
bivariate marginal densities.

Composite likelihood estimation methods form a subset of a more
general class of methods based on {\em proper scoring rules\/},
estimation being conducted by minimising the empirical score over
distributions in the model \citep{Musio2,Dawid:2014}.  Some important
proper scoring rules are the log-score, $S(x,q)=-\log{q(x)}$
\citep{Good}, which recovers the full (negative log) likelihood, and
the Brier score \citep{Brier:1950}.  A particularly interesting
special case, which entirely avoids the need to compute the
normalizing constant, is the {\em score matching\/} method of
\citet{Hyvarinen}, which is based on minimizing the following
objective function:
\begin{equation}\label{hiv}
  S(x,Q)=\Delta \ln{ q(x)} + \frac{1}{2}|| \nabla \ln{ q(x)}||^2,
\end{equation}
where $q(\cdot)$ is the density function of a distribution $Q$
proposed for a random variable $X$, and $x$ is the realized value of
$x$.  In \eqref{hiv}, $\nabla$ denotes the gradient operator, and
$\Delta$ the Laplacian operator, with respect to $x$.  This assumes
that the random variable $X$ is continuous-valued and defined over the
entire $\Real^k$ supplied with the standard norm $||\cdot||$, and that
$q$ is differentiable over $\Real^k$.  The score matching technique
was subsequently generalised to the case of a Riemannian manifold
\citep{Dawid:2005}, to the case of a non-negative real domain
$\Real^k_+$ or $\{\Real_+\cup 0\}^k$, and for binary variables
\citep{Hyvarinen:2007}.

The objective function in \eqref{hiv}, the ``Hyv\"arinen scoring
rule'', is a {\em 2-local homogeneous proper scoring rule\/}: see
\citet{Parry:2012}.  Inference performed using any homogeneous scoring
rule does not require the knowledge of the normalizing constant of the
distribution, since the value of the score is unaffected by applying a
positive scale factor to the density $q$.  Works considering
estimation based on the Hyv\"arinen score include
\citet{Musio1,Musio2,Forbes:2013}.  In a full natural exponential
family, score matching delivers a linear estimating equation, which
could be used as a starting point of iterative methods as in the {\tt
  R} package {\tt gRc} for Gaussian graphical model with symmetries
\citep{Forbes:2013,Lauritzen:2007}.

\indent The principal concern of this work is to investigate and
compare the behaviours of the estimators obtained from the Hyv\"arinen
score, the pairwise likelihood, and the full likelihood.  We confine
our attention to two different model settings: autoregressive and
moving average processes.  The loss in efficiency in using pairwise
likelihood methods may be slight in the former case, or very large, in
the latter \citep{Davis,Jin}.
\\
 
\indent The paper unfolds as follows.  Section~2 introduces basic
notions on scoring rules.  Section~3 describes estimation procedures
for first order autoregressive and moving average processes.
Section~4 summarizes the results of the simulation studies conducted.
We conclude in Section~5.

\section{Scoring rules}
A {\em scoring rule\/} is a loss function designed to measure the
quality of a proposed probability distribution $Q$, for a random
variable $X$ taking values in $\chi$, in view of the outcome $x$ of
$X$.  Specifically, if a forecaster quotes a predictive distribution
$Q$ for $X$ and the event $X=x$ realizes, then the loss will be
$S(x,Q)$.  The expected value of $S(X,Q)$ when $X$ has distribution
$P$ is denoted by $S(P,Q)$.
The scoring rule $S$ is {\em proper\/} (relative to the class of
distributions $\mathcal{P}$) if
\begin{equation}
  \label{eq:prop}
  S(P,Q)\geq S(P,P),\quad \textrm{for all}\;\;  P,\,Q\in\mathcal{P}.
\end{equation}
It is {\em strictly proper\/} if equality in \eqref{eq:prop} obtains
only when $Q = P$.

\subsection{Estimation} 
Let $(x_1, .  .  .  , x_{\nu})$ be independent realizations of a
random variable $X$, having distribution $P_{\theta}$ depending on an
unkown parameter $\theta\in\Theta$, where $\Theta$ is an open subset
of $\Real^k$.  Given a proper scoring rule $S$, let $S(x, \theta)$
denote $S(x, P_\theta)$.  Inference for the parameter $\theta$ may be
performed by minimising the {\em total empirical score\/},
\begin{equation*}
  S(\theta)=\sum_{i=1}^{\nu} S(x_i,\theta),
\end{equation*}
resulting in the {\em minimum score estimator\/},
\begin{equation*}
  \widehat{\theta}_S = \arg\min_{\theta}S(\theta).
\end{equation*}
Under broad regularity conditions on the model (see {\em e.g.\/}
\citet{B-N}), $\widehat{\theta}_S$ is a root of the {\em score
  equation\/}:
\begin{equation*}\label{score}
  s(\theta) :=\sum_{i=1}^{\nu} s(x_i, \theta)=0, 
\end{equation*}  
where $s(x, \theta)$ denotes the gradient vector of $S(x,\theta)$ with
respect to $\theta$: $s(x, \theta)= \nabla S(x,\theta)$.  The score
equation is an unbiased estimating equation \citep{Dawid:2005}.  When
$S$ is the log-score, the minimum score estimator coincides with the
maximum likelihood estimator.

From the general theory of unbiased estimating functions, under broad
regularity conditions on the model the minimum score estimate
$\widehat{\theta}_S$ is asymptotically consistent and normally
distributed:
$$\widehat{\theta}_S \sim N_{k}(\theta, \left\{\nu G(\theta)\right\}^{-1}),$$
where $G(\theta)$ denotes the {\em Godambe information matrix\/} (see
\citet{Dawid:2014,Musio2}):
\begin{equation*}\label{godambe}
  G(\theta):=K(\theta)J(\theta)^{-1}K(\theta),
\end{equation*}
where $J(\theta)=E\left[ s(X,\theta)s(X,\theta)^T\right]$ is the {\em
  variability matrix\/}, and $K(\theta)=E\left[\nabla
  s(X,\theta)^T\right]$ is the {\em sensitivity matrix\/}; in contrast
to the case for full likelihood, $J$ and $K$ are different in general.
It is possible to define test statistics, analogous to those based on
the full likelihood, starting from an arbitrary proper scoring rules:
{\em e.g.\/} scoring rule Wald-type, scoring rule score-type and
scoring rule ratio statistics \citep{Dawid:2014}.

\subsection{Standard errors} Estimation of the matrix $J(\theta)$, and
(to a somewhat lesser extent) of the matrix $K(\theta)$, is not an
easy task.  Here, we review the methods we use to estimate these two
matrices in the simulation studies.
\\
Let $(y_{1},\ldots , y_{\nu})$ be independent observations from a
$T$-dimensional distribution $P_{\theta}$.  If $\nu$ is quite large,
empirical estimation of the two matrices could be done as
\begin{align*}
  \widehat{J}&=\frac{1}{\nu}\sum_{i=1}^{\nu}s(y_{i},\widehat{\theta}_S)s(y_{i},\widehat{\theta}_S)^T,
  & \widehat{K}&=\frac{1}{\nu}\sum_{i=1}^{\nu}\frac{\partial
    s(y_{i},\theta)}{\partial \theta}\Big{|}_{\theta=\widehat{\theta}_S},
\end{align*}
with $y_{i} = (y_{1i},\ldots, y_{Ti})$.
\\
When it is possible to simulate directly from the complete model, the
two matrices could be estimated by recovering to their Monte Carlo
estimates, i.e.\
\begin{align*}
  \widehat{J}&=\frac{1}{B}\sum_{b=1}^{B}
  s(y^{(b)},\widehat{\theta}_S)s(y^{(b)},\widehat{\theta}_S)^{T}, &
  \widehat{K}&=\frac{1}{B}\sum_{b=1}^{B} \frac{\partial
    s(y^{(b)},\theta)}{{\partial
      \theta}}\Big{|}_{\theta=\widehat{\theta}_S},
\end{align*}
where $y^{(1)}, .  .  .  , y^{(B)}$ are $B$ independent realizations
from the model obtained by assuming $\widehat{\theta}_S$ as the true
parameter value.
\\
\indent We refer to \citet{Varin:2008} and \citet{Varin:2011} for a
detailed account on the estimation of the two matrices under the
composite likelihood setting.  \citet{Cattelan} compare the
performances of the composite likelihood based statistics (Wald-type,
score-type, and some adjustments of the composite likelihood ratio)
obtained by estimating $K$ and $J$ empirically with the ones produced
by using Monte Carlo simulation of the two matrices.

\section{The models}\label{models}
This section will be devoted to two examples both dealing with
multivariate normal distributions: the first order autoregressive and
moving average models which are two simple examples of linear time
series models.  They are chosen so that we can calculate both the full
and pairwise likelihood estimators.

\subsection{First order autoregressive models}\label{ar1}
The stationary univariate autoregressive process of order $1$, denoted
by $AR(1)$, is defined by
\begin{equation*}
  y_{t}-\mu= \phi(y_{t-1} -\mu) +z_{t},\quad \textrm{with}\quad t=2,\dots,T.
\end{equation*}
where $(z_{t})$ is Gaussian white noise process with mean $0$ and
variance $\sigma^2$, independent of the initial random variable $y_1$
which is a Gaussian random variable with mean $\mu$ and variance
${\sigma^2}/(1-\phi^2)$.  Here $\phi$, with $|\phi|<1$, is the {\em
  autoregressive parameter\/}.  Then $y_{1},\ldots,y_{T}$ are jointly
normal with mean vector $\mu1_T$ (where $1_T$ is the $T$-dimensional
unit vector), and covariance matrix $\Psi$ having components
$\psi_{lm}=\sigma^2{\phi^{|l-m|}}/{(1-\phi^2)}$ ($l, m = 1,\ldots,
T$).
\\

The full log-likelihood function for the unknown parameter $\theta =
(\mu, \sigma^2, \phi)$, based on data $y=(y_1,\ldots,y_T)$, is (see
for example \citet{Pace}): {\small
  \begin{eqnarray*}
    l(\theta)&=&-\frac{1}{2\sigma^2}\left\lbrace\sum_{t=1}^T (y_{t}-\mu)^2+\phi^2\sum_{t=2}^{T-1} (y_{t}-\mu)^2-2\phi\sum_{t=2}^T (y_{t}-\mu)(y_{t-1}-\mu)\right\rbrace\nonumber
    \\
    &&{}-\frac{T}{2}\log{\sigma^2}+\frac{1}{2} \log{(1-\phi^2)}.
  \end{eqnarray*}
}

As in \citet{Davis}, we shall consider the {\em consecutive pairwise
  likelihood\/}, rather than the complete pairwise likelihood, since
in the time series considered dependence decreases in time, so that
adjacent observations are more closely related than the others.
Since, for $t=2,\ldots,T$, $(y_t,y_{t-1})$ has a bivariate Gaussian
distribution, with common mean $\mu$ and variance
${\sigma^2}/(1-\phi^2)$, and covariance ${\sigma^2\phi}/(1-\phi^2)$,
the consecutive pairwise log-likelihood for $\theta = (\mu, \sigma^2,
\phi)$ is (see \citet{Pace}) {\small
  \begin{eqnarray*}
    pl(\theta)&=&-\frac{1}{2\sigma^2}\left\lbrace \sum_{t=2}^T(y_t-\mu)^2+\sum_{t=2}^T(y_{t-1}-\mu)^2-2\phi\sum_{t=2}^T(y_t-\mu)(y_{t-1}-\mu)\right\rbrace\nonumber
    \\
    &&{}-(T-1)\log{\sigma^2}+\frac{(T-1)}{2}\log(1-\phi^2).
  \end{eqnarray*}
}
\noindent When it is known that $\mu=0$, the pairwise likelihood
estimator, when both $\phi$ and $\sigma^2$ are of interest, has
components 
\begin{eqnarray*}
  \widehat{\phi}_p &=& 2\left(\frac{\sum_{t=2}^{T}y_ty_{t-1}}{\sum_{t=2}^{T} y_t^2+y_{t-1}^2
    }\right)\\
  \widehat{\sigma}^2_p &=&\left(
    \frac{\sum_{t=2}^{T}y_t^2+y_{t-1}^2}{2(T-1)}\right)^2(1-\widehat{\phi}_p^2).
\end{eqnarray*}
Note that $\widehat{\phi}_p$ is the Yule-Walker estimator \citep{Davis}.
\\

By using basic differentiation rules, it is easy to find the
Hyv\"arinen score for the model: {\small
  \begin{eqnarray*}\label{hivuni}
    H(y,\theta)&=&\frac{1}{2\sigma^4} \sum_{t=2}^{T-1} \left[(1+\phi^2) (y_{t}-\mu)-\phi(y_{t-1}+y_{t+1}-2\mu)\right]^2-\frac{2+(T-2)(1+\phi^2)}{\sigma^2}\nonumber
    \\
    &&{}+\frac{ \left\{y_{d}-\mu-\phi(y_{T-1}-\mu) \right\}^2}{2\sigma^4}  +\frac{\left\{y_{1}-\mu-\phi(y_{2}-\mu) \right\}^2}{2\sigma^4}.
  \end{eqnarray*}
}

The minimum score estimate of $\theta$, $\widehat{\theta}_H$, can be found
by minimising the Hyv\"arinen score in the above equation.

\subsection{First order moving average models}
The univariate moving average process of order $1$, denoted by
$MA(1)$, is defined by the equation
\begin{equation*}
  y_{t}-\mu= \alpha z_{t-1} +z_{t},\quad \quad (t=1,\dots,T),
\end{equation*}
where $|\alpha|<1$ and $z_0,\ldots,z_T$ are independent Gaussian
random variables with $0$ mean and variance $\sigma^2$.  Then the
random variables $y_{1},\ldots,y_{T}$ are jointly normal, each having
mean $\mu$ and variance $\sigma^2(1+\alpha^2)$.  The variables $y_t$
and $y_{t+k}$ are independent for $|k| > 1$, while $y_t$ and $y_{t+1}$
have covariance $\sigma^2\alpha$ ($t= 1,\ldots, T-1$).  Hence, the
covariance matrix $\Omega$ of $y=(y_1,y_2,\ldots,y_T)$ has components
$\omega_{ss}=\sigma^2(1+\alpha^2)$, $\omega_{st}=\sigma^2\alpha$ if
$|s-t|=1$, $\omega_{st}=0$ otherwise.
\\

Let $\theta = (\mu, \sigma^2,\alpha)$ be the vector of model
parameters, dropping all constant terms, the full log-likelihood
function of a single series is (see for instance
\citet[pag.128]{Hamilton})
\begin{eqnarray*}{\label{lik}}
  l(\theta)&=&-\frac{1}{2}\log{|\Omega|}-\frac{1}{2}(y-\mu)\Omega^{-1}(y-\mu)^T.
\end{eqnarray*}
The maximum likelihood estimator $\widehat{\theta}$ can be found by
maximizing numerically the above objective function.
\\

As before we consider the consecutive pairwise likelihood.  For
$t=2,\ldots,T$, the pair $(y_t,y_{t-1})$ has a bivariate Gaussian
density, in which the two components have both mean $\mu$ and variance
$\sigma^2(1+\alpha^2)$, and have covariance $\sigma^2\alpha$.  The
pairwise likelihood for contiguous pairs of observations of a single
series is thus
{\small
  \begin{eqnarray*}
    pl(\theta)&=&-\frac{1}{2\sigma^2} \sum_{t=2}^{T}\frac{\left\{(y_{t}-\mu)^2+(y_{t-1}-\mu)^2\right\}\left( 1+\alpha^2\right)-2(y_{t}-\mu)(y_{t-1}-\mu) \alpha}{1+\alpha^2+\alpha^4}
    \\
    &&{}-\frac{(T-1)}{2} \log{(1+\alpha^2+\alpha^4)}-(T-1)\log{\sigma^2}.  
  \end{eqnarray*}
} \indent The pairwise likelihood estimator $\widehat{\theta}_p$ can be
found by maximizing numerically the pairwise log-likelihood function.
\\

By using basic differentiation rules, it is easy to find the
Hyv\"arinen score based on variables $(y_1,y_2,\ldots,y_T)$: {\small
  \begin{eqnarray}\label{hivunima}
    H(y,\theta)&=&-\sum_{i=1}^T\omega^{ii}+\frac{1}{2}\sum_{i=1}^T\left\{ \sum_{t=1}^T\omega^{it}(y_t-\mu)\right\} ^2,
  \end{eqnarray}
} where $\omega^{ij}$ denotes the $(i,j)$ element of the inverse of
the matrix $\Omega$.

\subsection{$\nu$ independent series}
In the remainder of this paper we consider $\nu$ independent series of
length $T$.  We assume that $T$ is fixed while $\nu$ increases to
infinity.  We also specialise to the case that the common mean $\mu$
and variance $\sigma^2$ are known; without loss of generality we shall
assume $\mu=0$, $\sigma^2=1$.
\\

So consider now $\nu$ independent and identically distributed first
order autoregressive processes $Y_1, \ldots ,Y_{\nu}$, having
autoregressive parameter $\phi$.  Let the $(\nu\times T)$ random
matrix $Y$ have the vector $Y_i$ as its $i$th row: thus each row of
$Y$ is independent of the others, and has the $T$-variate normal
distribution with mean-vector $0$ and variance covariance matrix
$\Psi$ say.  An estimating function for the parameter $\phi$ can be
obtained by summing the $\nu$ individual Hyv\"arinen scores, or $\nu$
score equations, or $\nu$ pairwise score equations.  But we can also
take into consideration the fact that the sum-of-squares-and-products
matrix $S=Y^TY$ is a sufficient statistic for the multivariate normal
model, having the Wishart distribution with $\nu$ degrees of freedom
and scale matrix $\Psi$.  Then inference for the parameter $\phi$ can
be performed by resorting to the Hyv\"arinen score based directly on
the Wishart model.  T The same approach can be taken if we have $\nu$
independent first order moving average processes with the same moving
average parameter $\alpha$: Dawid and Musio (2014) apply this method
to a similar, but non-stationary, model having a tridiagonal
covariance matrix.

Assuming $\nu\geq T$ so that the joint distribution of the upper
triangle $\left(s_{ij}: 1\leq i\leq j\leq T\right)$ of the
sum-of-squares-and-products random matrix $S$ (which has a Wishart
distribution with parameters $\nu $ and $\Lambda$) has a density, and
taking into consideration all of the properties of the derivatives of
traces and determinants, it can be shown that the Hyv\"arinen score
based on this joint density is
{\small
  \begin{equation}\label{HS}
    HW(S,\Lambda)=-\frac{(\nu-T-1)}{2}\sum_{i=1}^{T}(s^{ii})^2+\frac{1}{2}\sum_{i,j=1}^{T}\left\{\frac{(\nu-T-1)}{2}s^{ji}-\frac{1}{2}\lambda^{ji}\right\}^2,
  \end{equation}
}
\noindent where $s^{ij}$, $\lambda^{ij}$ are the elements of the
inverse matrices $S^{-1}$ and $\Lambda^{-1}$, respectively.
\\
If the scale matrix $\Lambda$ is modelled in terms of a scalar
parameter $\lambda$ (where $\lambda=\phi$ or $\alpha$ in our models),
the associated estimate $\widehat{\lambda}_{HW}$ is now found by
minimising $HW(S,\Lambda)$ with respect to $\lambda$.

However, for both our models, the Godambe Information needed to
estimate the standard error of $\widehat{\lambda}_{HW}$ is not easy to
derive analytically.  The derivative of $HW(S,\Lambda)$ with respect
to $\lambda$ is
\begin{equation}\label{der}
  HW_{\lambda}(S,\Lambda)=-\frac{1}{2}\sum_{i,j=1}^{T}\left\{\frac{(\nu-T-1)}{2}s^{ji}-\frac{1}{2}\lambda^{ji}\right\}\frac{\partial\lambda^{ji}}{\partial\lambda},
\end{equation}
\noindent and $E\left\{HW_{\lambda}(S,\Lambda)\right\}=0$ since
$E\left( s^{ij}\right)={\lambda^{ij}}/(\nu-T-1) $ (see
\citet[p.~257]{Kollo}).  Moreover, $K(\lambda)=E\left\{
  HW_{\lambda\lambda}(S,\Lambda)\right\}=\frac{1}{4}\sum_{i,j=1}^{T}\left({\partial\lambda^{ji}}/{\partial\lambda}\right)^2$.
Given the simple form of the inverse of the matrix $\Psi$ in the
$AR(1)$ model, a tridiagonal matrix with elements above and below the
main diagonal equal to $-\phi$, and all diagonal elements equals to
$(1+\phi^2)$ except for the elements $\psi^{11}$ and $\psi^{TT}$ which
are equal to $1$ (see for instance \citet{Davison}), the function $K$
reduces to
\begin{equation}\label{Kar}
  K(\phi)=\frac{T-1+2\phi^2(T-2)}{2}.
\end{equation}
The function $K$ for the $MA(1)$ model entails more lengthy
calculations since the elements of the inverse of the matrix $\Omega$
are (see for example \citet{Shaman})
\begin{equation}
  \omega^{ij}=(-\alpha)^{j-i}\frac{\left(1+\alpha^2+\ldots+\alpha^{2(i-1)}\right) \left( 1+\alpha^2+\ldots+\alpha^{2(T-j)}\right) }{\left( 1+\alpha^2+\ldots+\alpha^{2T}\right) },\quad j\geq i.
\end{equation}
The derivation of the function $J(\lambda)$, which after taking
account of the square of \eqref{der} reduces to
\begin{equation}
  J(\lambda)=\frac{(\nu-T-1)^2}{16}\sum_{i,j,k,l=1}^{T}\left( \frac{\partial\lambda^{ji}}{\partial\lambda}\right) ^2\mbox{cov}\left( s^{ji},s^{kl}\right),
\end{equation} 
involves calculations requiring the covariance matrix of the random
matrix $S^{-1}$, which has an Inverse Wishart distribution with scale
matrix $\Lambda^{-1}$: see \citet{Rosen} for details on the components
of the covariance matrix.
\\
It should be pointed out that this approach can not be used if we have
a single time series of length $T$ with $T$ increasing to $\infty$,
since for non-singularity of the Wishart distribution we need to
assume $\nu\geq T$.

\section{Simulation studies}
We designed two simulation studies to assess and compare the
behaviours of the estimators found by using the Hyv\"arinen scoring
rule and the full and pairwise maximum likelihood estimators.  In
Simulation~1 we assume a first order autoregressive model, while in
Simulation~2 we consider a first order moving average process.
Various parameter settings are considered in both simulation studies.
All calculations have been done in the statistical computing
environment {\tt R} \citep{R:2013}.  In both simulations, $1000$
replicates are generated of $\nu = 200$ processes of length $T=50$.
(Similar results, not reported here, were obtained with $\nu$
increased to $300$.)
\\

In Simulation~1, the values of the model parameters are $\mu=0$ and
$\sigma=1$, with the autoregressive parameter
$\phi\in\{-0.9,-0.8,\ldots,0.8,0.9\}$.  Results are summarized in
Table~\ref{tab1}, which reports average estimates of the
autoregressive parameter $\phi$ using the full likelihood
($\widehat{\phi}$), the pairwise likelihood ($\widehat{\phi}_{p}$), the sum of
$\nu$ Hyv\"arinen scores ($\widehat{\phi}_{H}$), and the Hyv\"arinen score
based on the Wishart model ($\widehat{\phi}_{HW}$).  Moreover, it provides
the asymptotic standard deviations ($sd$) and the relative asymptotic
efficiency ($ARE$) with respect to the maximum likelihood estimator
$\widehat{\phi}$, i.e.\ the ratio between the Fisher information and the
Godambe function.
\\

\begin{table}[htb]
  \caption{Estimated mean  ($Est.$),  asymptotic standard deviation ($sd$), and asymptotic relative efficiency ($ARE$) of estimators  of the parameter $\phi$ in the $AR(1)$ model, for $\nu=200$, $T=50$, and varying values of $\phi$.}
  \label{tab1}       
  \centering
  \vspace{0.2cm}
  \scalebox{0.76}{
    \renewcommand{\arraystretch}{1.5}
    \setlength{\tabcolsep}{3.1pt}
    \begin{tabular}{rlrllrlllrlllrlll}
      \hline
      \multicolumn{1}{l}{} & \multicolumn{1}{l}{} & \multicolumn{2}{c}{$\widehat{\phi}$} &  \multicolumn{1}{l}{} & \multicolumn{3}{c}{$\widehat{\phi}_p$} &  \multicolumn{1}{l}{} & \multicolumn{3}{c}{$\widehat{\phi}_H$} &  \multicolumn{1}{l}{} & \multicolumn{3}{c}{$\widehat{\phi}_{HW}$}\\
      \cline{3-4} \cline{6-8} \cline{10-12} \cline{14-16} 
      \multicolumn{1}{c}{$\phi$} &   \multicolumn{1}{l}{}  & \multicolumn{1}{c}{$Est.$} & \multicolumn{1}{c}{$sd$} &   \multicolumn{1}{l}{} & \multicolumn{1}{c}{$Est.$} & \multicolumn{1}{c}{$sd$} & \multicolumn{1}{c}{$ARE$} &  \multicolumn{1}{l}{} &  \multicolumn{1}{c}{$Est.$} & \multicolumn{1}{c}{$sd$} & \multicolumn{1}{c}{$ARE$} &  \multicolumn{1}{l}{}  &  \multicolumn{1}{c}{$Est.$} & \multicolumn{1}{c}{$sd$} & \multicolumn{1}{c}{$ARE$}\\
      \hline
      $-0.9$ & $ $ & $-0.8997$ & $0.0041$ & $ $ & $-0.8997$ & $0.0045$ & $0.8625$ & $ $ & $-0.9008$ & $0.0150$ & $0.0738$ & $ $ & $-0.9004$ & $0.0244$ & $0.0278$\\
      $-0.8$ & $ $ & $-0.8000$ & $0.0059$ & $ $ & $-0.7999$ & $0.0064$ & $0.8340$ & $ $ & $-0.8007$ & $0.0146$ & $0.1602$ & $ $ & $-0.8007$ & $0.0236$ & $0.0613$\\
      $-0.7$ & $ $ & $-0.7002$ & $0.0071$ & $ $ & $-0.7001$ & $0.0079$ & $0.8087$ & $ $ & $-0.7007$ & $0.0139$ & $0.2599 $ & $ $ & $-0.7005$ & $0.0226$ & $0.0979$\\
      $-0.6$ & $ $ &  $-0.6002$ & $0.0080$ & $ $ & $-0.6002$ & $ 0.0089$ &  $0.7986$ & $ $ &  
      $-0.6008$ & $0.0130$ & $0.3794$ & $ $ & $-0.6008$ & $0.0216$ & $0.1367$ \\
      $-0.5$ & $ $ & $-0.5001$ & $0.0087$ & $ $ & $-0.4999$ & $0.0097$ & $0.8069$ & $ $ &  
      $-0.5009$ & $0.0122$ & $0.5060$ & $ $ & $ -0.5011$ & $0.0202$ & $0.1853$ \\
      $-0.4$ & $ $ & $-0.4002$  & $0.0092$ & $ $ & $-0.4000$ & $0.0101$ & $0.8351$ & $ $ &
      $-0.4006$ & $ 0.0115$ & $0.6466$ & $ $  & $-0.4001$ &  $0.0184$ & $0.2505$ \\
      $-0.3$ & $ $ & $-0.2997$ & $0.0096$ & $ $ & $-0.2997$ & $0.0102$ & $0.8808$ & $ $ & $-0.2998$ & $0.0109$ &  $0.7773$ & $ $ & $-0.2995$ & $0.0164$ & $0.3438$\\
      $-0.2$ & $ $ & $-0.2003$ &  $0.0099$ & $ $ &  $-0.2002$ & $0.0102$ & $0.9347$ & $ $ & $-0.2005$ &  $0.0104$ & $0.8991$ & $ $ & $-0.2007$ & $0.0143$ & $0.4780$\\
      $-0.1$ & $ $ & $-0.0997$ & $0.0100$ & $ $ & $-0.0997$ & $0.0101$ & $0.9813$ & $ $ & $-0.0997$ & $0.0102$ & $0.9776$ & $ $ & $-0.0999$ & $0.0125$ & $0.6493$\\
      $0$ & $ $ & $0.0002$ & $0.0101$ & $ $ & $0.0002$ & $0.0101$ & $0.9998$ & $ $ & $0.0002$ & $0.0101$ & $1.0077$ & $ $ & $0.0003$ & $0.0117$ & $0.7401$\\
      $0.1$ & $ $ & $0.1005$ & $0.0100$ & $ $ & $0.1005$ & $0.0101$ & $0.9810$ & $ $ & $0.1005$ & $0.0101$ & $0.9810$ & $ $ & $0.1007$ & $0.0125$ & $0.6506$ \\
      $0.2$ & $ $ & $0.1997$ & $0.0099$ & $ $ & $0.1997$ & $0.0102$ & $0.9350$ $ $ & $ $ &  $0.1998$ 
      & $0.0104$ & $0.8980$ & $ $ & $0.1995$ & $0.0143$ &  $0.4802$  \\
      $0.3$ & $ $ & $0.2997$ & $0.0096$ & $ $ & $0.2997$ & $0.0102$ & $0.8808$ & $ $ & $0.2998$ & $0.0109$ & $0.7774$ & $ $ & $0.2995$ & $0.0164$ & $0.3433$\\                                              
      $0.4$ & $ $ & $0.3993$ & $0.0092$ & $ $ & $0.3993$ & $0.0101$ & $0.8355$ & $ $ & $0.3997$ & $0.0115$ & $0.6451$ & $ $ & $0.3995$ & $0.0184$ & $0.2506$\\
      $0.5$ & $ $ & $0.5002$ & $0.0087$ & $ $ & $0.5003$ & $0.0097$ & $0.8071$ & $ $ & $0.5006$ & $0.0122$ & $0.5077$ & $ $ & $0.5004$ & $0.0201$ & $0.1867$\\
      $0.6$ & $ $ &  $0.5997$ & $0.0080$ & $ $ &  $0.5997$ & $0.0089$ & $0.7985$ & $ $ & $0.5998$ & $0.0130$ & $0.3757$ & $ $ & $0.5990$ & $0.0215$ & $0.1376$\\
      $0.7$ & $ $ & $0.6992$ & $0.0071$ &  $ $ & $0.6992$ & $0.0079$ & $0.8087$ & $ $ & $0.6997$ & $0.0138$ & $0.2630$ & $ $ & $0.6993$ & $0.0227$ & $0.0977$\\
      $0.8$ & $ $ &  $0.8001$ & $0.0058$ & $ $ & $0.8001$ & $0.0064$ & $0.8343$ & $ $ & $0.8006$ & $0.0146$ & $0.1605$ & $ $ & $0.8002$ & $0.0235$ & $0.0618$\\
      $0.9$ & $ $ & $0.8998$ & $0.0041$ & $ $ & $0.8998$ & $0.0044$ & $0.8622$ & $ $ & $0.8999$ & $0.0150$ & $0.0734$ &$ $ & $0.8987$ & $0.0244$ & $ 0.0278$\\
      \hline
    \end{tabular}
  }
\end{table}

In Simulation~2, the values of the model parameters are $\mu=0$ and
$\sigma=1$, with the moving average parameter
$\alpha\in\{-0.9,-0.8,\ldots,0.8,0.9\}$.  Results are summarized in
Table~\ref{tab2}, which shows the estimates of the moving average
parameter $\alpha$ using the full likelihood ($\widehat{\alpha}$), the
pairwise likelihood ($\widehat{\alpha}_{p}$), the sum of $\nu$ Hyv\"arinen
scores ($\widehat{\alpha}_{H}$), and the Hyv\"arinen score based on the
Wishart model ($\widehat{\alpha}_{HW}$) with the average of the associated
standard errors ($sd$) and the asymptotic relative efficiency with
respect to the maximum likelihood estimator $\widehat{\alpha}$ ($ARE$).
\\

\begin{table}[htb]
  \caption{Estimated mean  ($Est.$),  asymptotic standard deviation ($sd$), and asymptotic relative efficiency ($ARE$) of estimators  of the parameter $\alpha$ in the $MA(1)$ model, for $\nu=200$, $T=50$, and varying values of $\alpha$.}
  \vspace{0.2cm}
  \label{tab2}       
  \centering
  \scalebox{0.76}{
    \renewcommand{\arraystretch}{1.5}
    \setlength{\tabcolsep}{3.1pt}
    \begin{tabular}{rlrllrlllrlllrlll}
      \hline
      \multicolumn{1}{l}{} & \multicolumn{1}{l}{} & \multicolumn{2}{c}{$\widehat{\alpha}$} &  \multicolumn{1}{l}{} & \multicolumn{3}{c}{$\widehat{\alpha}_p$} &  \multicolumn{1}{l}{} & \multicolumn{3}{c}{$\widehat{\alpha}_H$} &  \multicolumn{1}{l}{} & \multicolumn{3}{c}{$\widehat{\alpha}_{HW}$}\\
      \cline{3-4} \cline{6-8} \cline{10-12} \cline{14-16} 
      \multicolumn{1}{c}{$\alpha$} &   \multicolumn{1}{c}{}  & \multicolumn{1}{c}{$Est.$} & \multicolumn{1}{c}{$sd$} &   \multicolumn{1}{c}{} & \multicolumn{1}{c}{$Est.$} & \multicolumn{1}{c}{$sd$} & \multicolumn{1}{c}{$ARE$} &  \multicolumn{1}{c}{} &  \multicolumn{1}{c}{$Est.$} & \multicolumn{1}{c}{$sd$} & \multicolumn{1}{c}{$ARE$} &  \multicolumn{1}{c}{}  &  \multicolumn{1}{c}{$Est.$} & \multicolumn{1}{c}{$sd$} & \multicolumn{1}{c}{$ARE$}\\
      \hline $-0.9$ & $ $ & $-0.8998$ & $0.0055$ & $ $ & $-0.8996$ & $0.0167$ & $0.1064$ & $ $ & $-0.8999$ & $0.0064$ & $0.7208$ & $ $ & $-0.8993$ & $0.0074$ & $0.5471$\\
$-0.8$ & $ $ & $-0.7997$ & $0.0066$ & $ $ & $-0.7996$ & $0.0176$ & $0.1390$ & $ $ & $-0.7998$ & $0.0075$ & $0.7566$ & $ $ & $-0.7992$ & $0.0091$ & $0.5177$\\
$-0.7$ & $ $ & $-0.6997$ & $0.0075$ & $ $ & $-0.6996$ & $0.0183$ & $0.1692$ & $ $ & $-0.6997$ & $0.0086$ & $0.7583$ & $ $ & $-0.6993$ & $0.0106$ & $0.5020$\\
$-0.6$ & $ $ & $-0.6004$ & $0.0083$ & $ $ & $-0.6005$ & $0.0182$ & $0.2080$ & $ $ & $-0.6007$ & $0.0095$ & $0.7553$ & $ $ & $-0.6003$ & $0.0119$ & $0.4878$\\
$-0.5$ & $ $ & $-0.5004$ & $0.0089$ & $ $ & $-0.4999$ & $0.0169$ & $0.2757$ & $ $ & $-0.5007$ & $0.0101$ & $0.7646$ & $ $ & $-0.5002$ & $0.0129$ & $0.4743$\\                                                                                                                                                                                   
$-0.4$ & $ $ & $-0.4000$ & $0.0093$ & $ $ & $-0.3997$ & $0.0148$ & $0.3984$ & $ $ & $-0.4003$ & $0.0104$ & $0.8038$ & $ $ & $-0.4001$ & $0.0136$ & $0.4713$\\
$-0.3$ & $ $ & $-0.3003$ & $0.0097$ & $ $ & $-0.3000$ & $0.0126$ & $0.5905$ & $ $ & $-0.3006$ & $0.0105$ & $0.8527$ & $ $ & $-0.3006$ & $0.0139$ & $0.4838$\\
$-0.2$ & $ $ & $-0.2000$ & $0.0099$ & $ $ & $-0.2002$ & $0.0111$ & $0.7926$ & $ $ & $-0.2001$ & $0.0104$ & $0.9119$ & $ $ & $-0.1999$ & $0.0135$ & $0.5408$\\
$-0.1$ & $ $ & $-0.1003$ & $0.0101$ & $ $ & $-0.1004$ & $0.0103$ & $0.9456$ & $ $ & $-0.1004$ & $0.0101$ & $0.9882$ & $ $ & $-0.1006$ & $0.0124$ & $0.6557$\\
$ 0  $ & $ $ & $ 0.0001$ & $0.0101$ & $ $ & $0.0001 $ & $0.0101$ & $1.0082$ & $ $ & $ 0.0001$ & $0.0101$ & $1.0101$ & $ $ & $ 0.0005$ & $0.0117$ & $0.7429$\\ 
$ 0.1$ & $ $ & $ 0.1000$ & $0.0101$ & $ $ & $0.1000 $ & $0.0103$ & $0.9526$ & $ $ & $ 0.1001$ & $0.0101$ & $0.9933$ & $ $ & $ 0.0997$ & $0.0124$ & $0.6554$\\
$ 0.2$ & $ $ & $ 0.2000$ & $0.0099$ & $ $ & $0.2000 $ & $0.0111$ & $0.7932$ & $ $ & $ 0.2000$ & $0.0104$ & $0.9171$ & $ $ & $ 0.1994$ & $0.0135$ & $0.5402$\\
$ 0.3$ & $ $ & $ 0.2994$ & $0.0097$ & $ $ & $0.2996 $ & $0.0126$ & $0.5853$ & $ $ & $ 0.2994$ & $0.0105$ & $0.8475$ & $ $ & $ 0.2992$ & $0.0139$ & $0.4835$\\
$ 0.4$ & $ $ & $ 0.4000$ & $0.0093$ & $ $ & $0.4006 $ & $0.0148$ & $0.3979$ & $ $ & $ 0.4000$ & $0.0105$ & $0.7938$ & $ $ & $ 0.3994$ & $0.0137$ & $0.4639$\\
$ 0.5$ & $ $ & $ 0.5002$ & $0.0089$ & $ $ & $0.5000 $ & $0.0169$ & $0.2760$ & $ $ & $ 0.5004$ & $0.0101$ & $0.7672$ & $ $ & $ 0.5000$ & $0.0129$ & $0.4721$\\
$ 0.6$ & $ $ & $ 0.6001$ & $0.0083$ & $ $ & $0.6000 $ & $0.0182$ & $0.2075$ & $ $ & $ 0.6001$ & $0.0095$ & $0.7643$ & $ $ & $ 0.5993$ & $0.0119$ & $0.4850$\\
$ 0.7$ & $ $ & $ 0.6999$ & $0.0075$ & $ $ & $0.6997 $ & $0.0182$ & $0.1707$ & $ $ & $ 0.6999$ & $0.0086$ & $0.7682$ & $ $ & $ 0.6996$ & $0.0106$ & $0.5047$\\
$ 0.8$ & $ $ & $ 0.7999$ & $0.0066$ & $ $ & $0.7997 $ & $0.0175$ & $0.1402$ & $ $ & $ 0.8000$ & $0.0075$ & $0.7639$ & $ $ & $ 0.7995$ & $0.0091$ & $0.5209$\\
$ 0.9$ & $ $ & $ 0.8999$ & $0.0055$ & $ $ & $0.8997 $ & $0.0167$ & $0.1072$ & $ $ & $ 0.9000$ & $0.0064$ & $0.7300$ & $ $ & $ 0.8995$ & $0.0074$ & $0.5504$\\
      \hline
    \end{tabular}
  }
\end{table}  

It should be noted that for the $MA(1)$ model no analytic expressions
for the derivatives of \eqref{hivunima} are available.  Numerical
evaluation of scoring rule derivatives has been carried out using the
{\tt R} package {\tt numDeriv}.
\\

\begin{figure}[htb]
  \centering \subfigure
  {\includegraphics[scale=0.4,keepaspectratio]{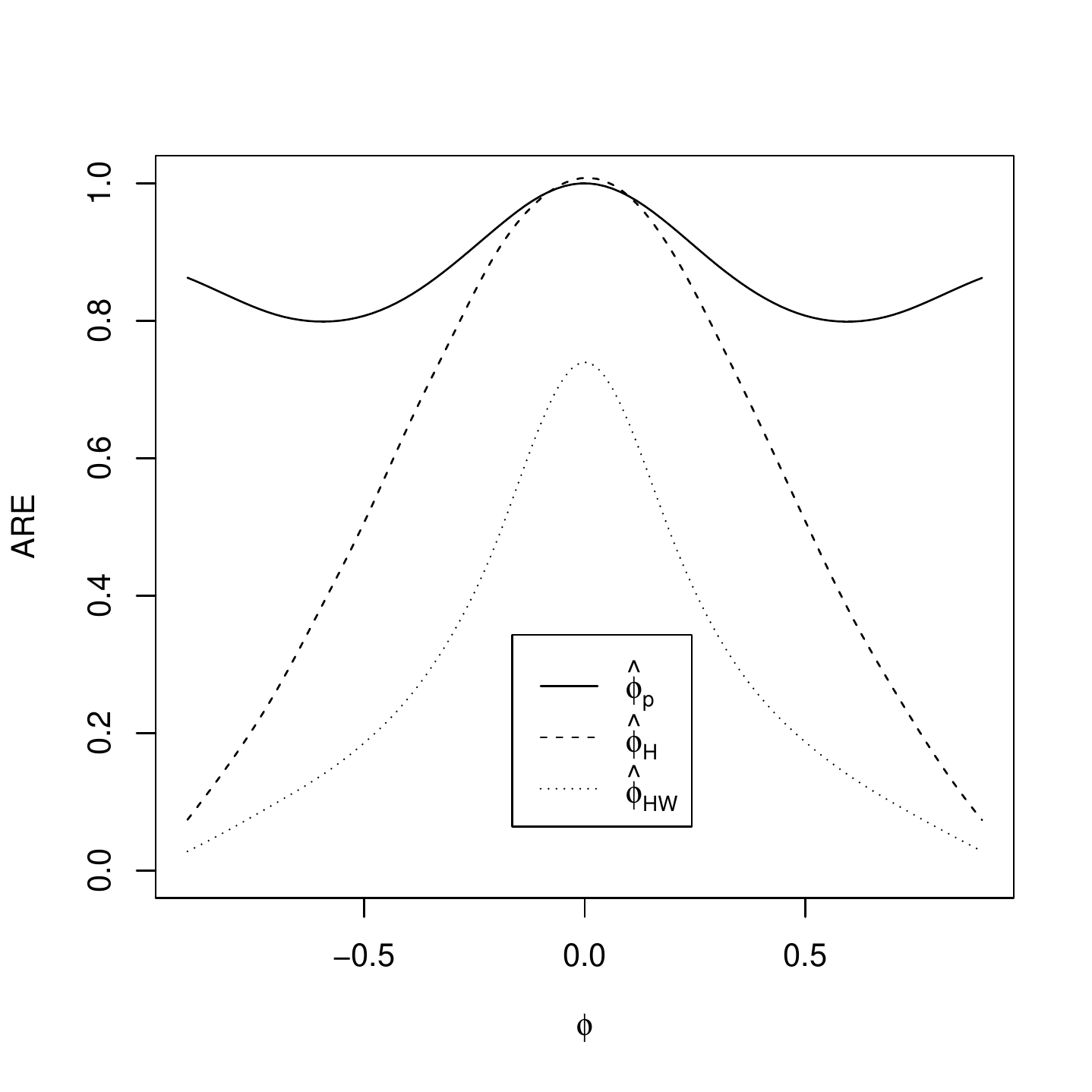}} \subfigure
  {\includegraphics[scale=0.4,keepaspectratio]{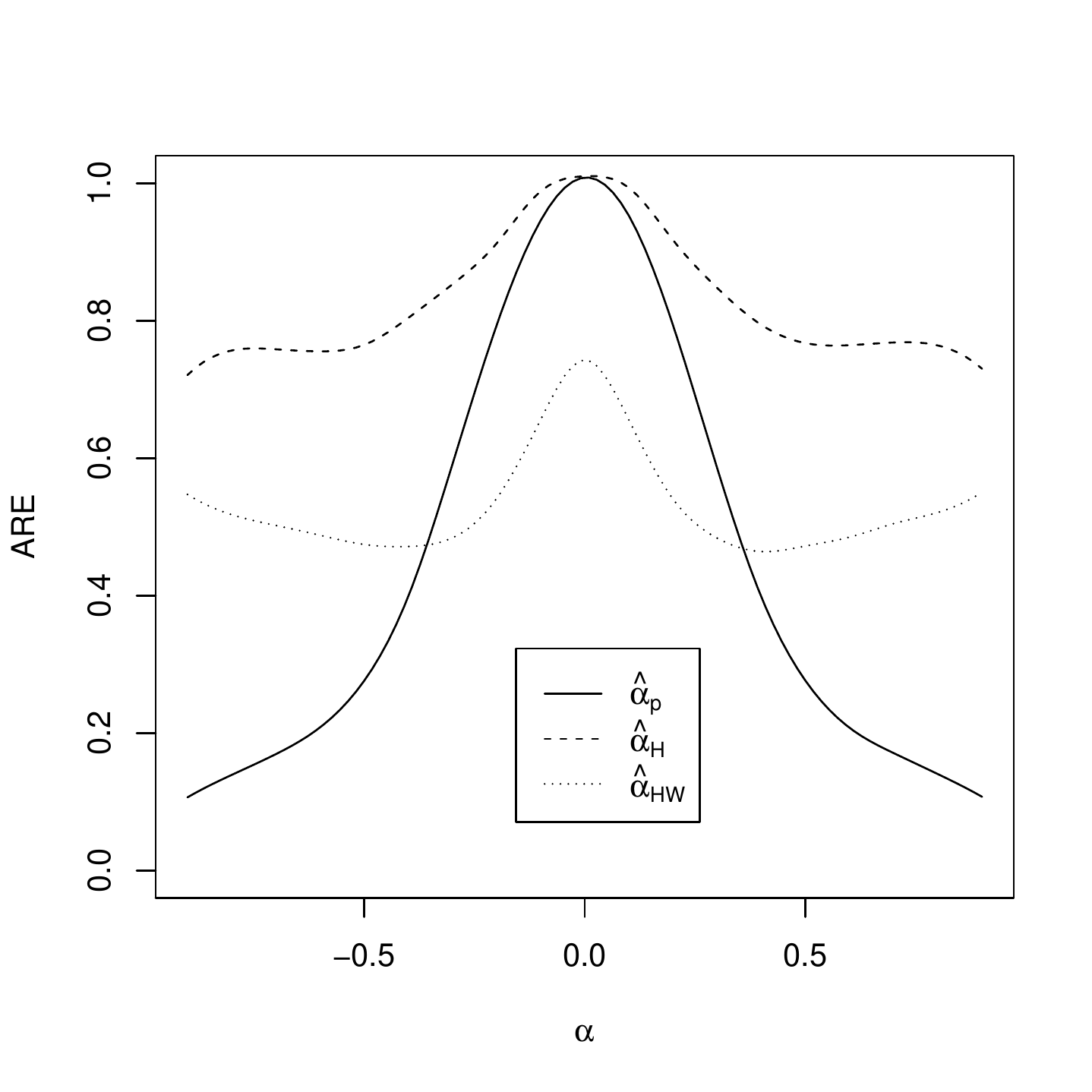}}
  \caption{Asymptotic relative efficiency of estimators for the
    $AR(1)$ model (left panel) and for the $MA(1)$ model (right
    panel).  Based on $1000$ replications of $\nu=200$ series of
    length $T =50$.}
  \label{fig1}
\end{figure}

\begin{figure}[htb]
  \centering \subfigure
  {\includegraphics[scale=0.4,keepaspectratio]{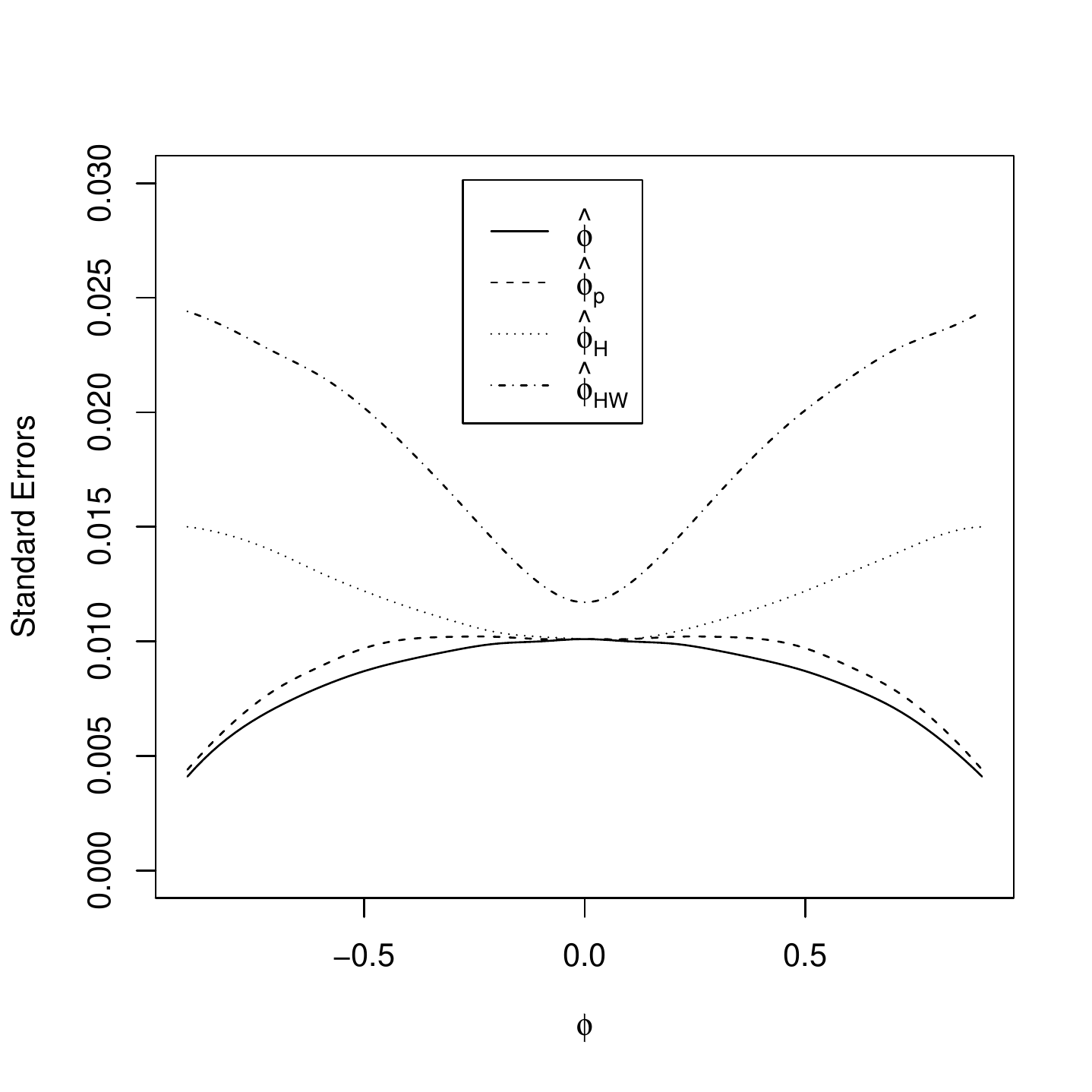}} \subfigure
  {\includegraphics[scale=0.4,keepaspectratio]{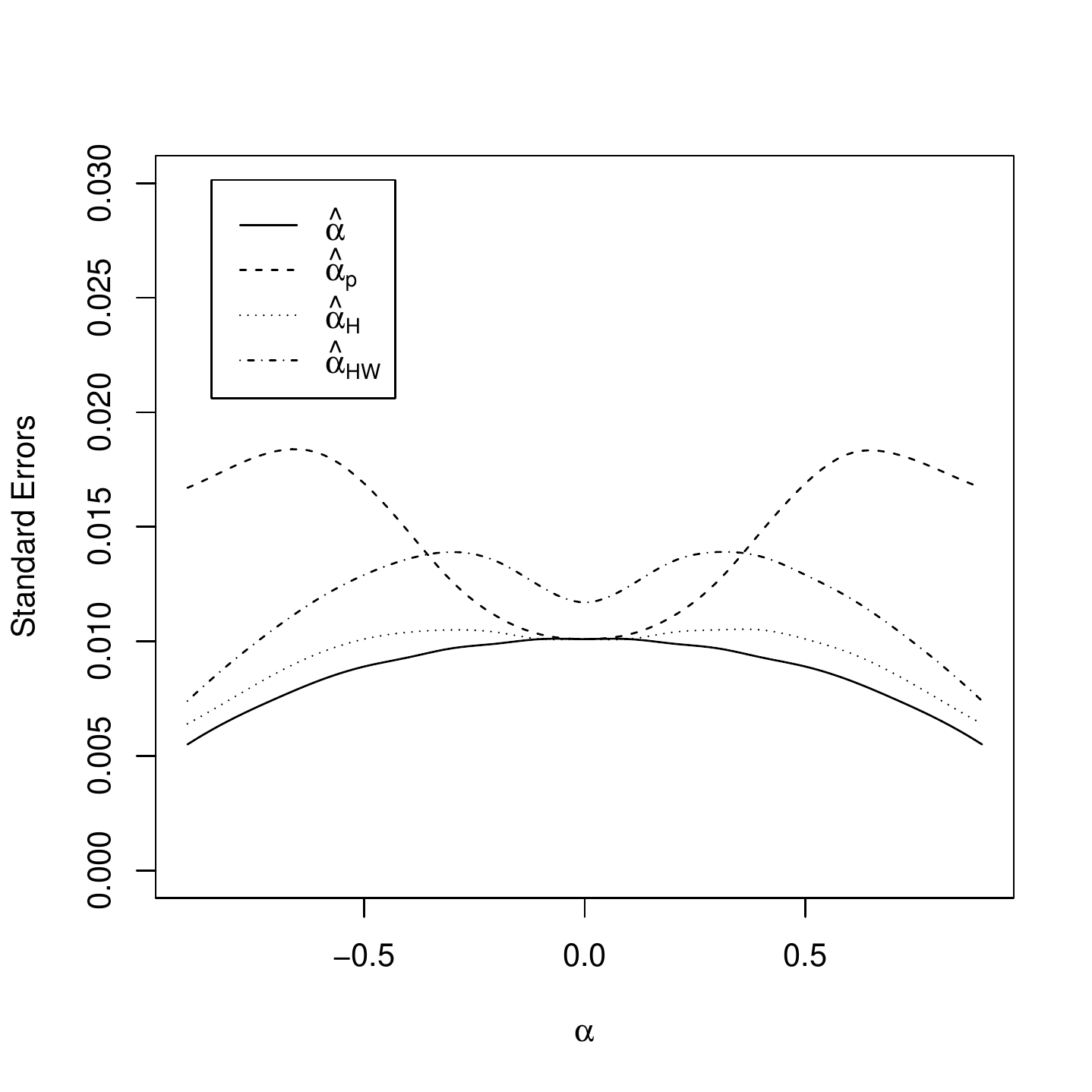}}
  \caption{Standard errors of estimators for the $AR(1)$ model (left
    panel), and for the $MA(1)$ model (right panel).  Based on $1000$
    replications of $\nu=200$ series of length $T =50$.}
  \label{fig2}
\end{figure}

The standard deviations of $\widehat{\phi}_{H}$ and $\widehat{\alpha}_{H}$ are
empirical estimates of the square root of the Godambe information
function, which is obtained by compounding the empirical estimates of
$J$ and $K$.  The standard deviations of the pairwise maximum
likelihood estimator and the maximum likelihood estimator are obtained
by using the analytic expressions (see \citet{Pace}) for the $AR(1)$
model and the empirical counterparts for the $MA(1)$ model.  The
Godambe information function of $\widehat{\phi}_{HW}$ and
$\widehat{\alpha}_{HW}$ are estimated by Monte Carlo simulations:
specifically, in the $AR(1)$ model we resort to analytic derivatives
of \eqref{HS} for the implementation of $J$ and to the analytical form
of $K$ in equation \eqref{Kar}; while in the $MA(1)$ model we use
numerical derivatives of \eqref{HS} for calculating both $K$ and $J$.
\\
\indent The left and right-hand panels of Figure~\ref{fig1} depict the
asymptotic relative efficiency as a function of $\phi$ for the $AR(1)$
model and as a function of $\alpha$ for the $MA(1)$ model for
$\nu=200$ and $T=50$, respectively.
\\
The left and right-hand panels of Figure~\ref{fig2} show the standard
errors as a function of $\phi$ for the $AR(1)$ model and as a function
of $\alpha$ for the $MA(1)$ model, for $\nu=200$ and $T=50$.
\\


\subsection{Discussion}
Results from Simulations~1 and 2 reveal that the estimators considered
produce estimates very close to the true values of the parameters.
However, results not shown here suggest that when the length $T$ of
the series is small the pairwise likelihood estimator performs worse
in terms of bias than the other estimators in both the models.  The
numerical results in Table~\ref{tab1} and in the left-hand panel of
Figure~\ref{fig1} suggest that $\widehat{\phi}_{H}$ and $\widehat{\phi}_{HW}$
do not have high efficiency as $\phi$ approaches $1$: in particular,
the asymptotic efficiency of $\widehat{\phi}_{HW}$ tends to $0$ for large
values of $|\phi|$. In contrast, under the same model setting, there
is only a modest loss of efficiency for the pairwise likelihood-based
estimator.  Simulation~2 shows that the univariate Hyv\"arinen
estimator $\widehat{\alpha}_{H}$ achieves the same efficiency as the $MLE$
in the $MA(1)$ model for values of the moving average parameter near
$0$; see Table~\ref{tab2} and the right-hand panel of
Figure~\ref{fig1}.  However, the loss in efficiency of the univariate
Hyv\"arinen estimator $\widehat{\alpha}_{H}$ is modest even when the
absolute value of the moving average parameter reaches $1$.  The
standard errors of the univariate and the multivariate Hyv\"arinen
estimators increase as $|\alpha|$ increases from $0$ to $0.3$ and
decrease as $|\alpha|$ increases from $0.3$ to $0.9$; see the
right-hand panel of Figure~\ref{fig2}.  In contrast, the pairwise
method shows very poor performances in terms of asymptotic relative
efficiency: the $ARE$ ranges from $1$ to $0.1$ as $|\alpha|$
increases.  These results are in agreement with the findings of
\cite{Davis} who focus on pairwise likelihood-based methods for linear
time series.

\section{Conclusions}
We have investiged the performance of two estimators based on the
Hyv\"arinen scoring rule, which can be regarded as a surrogate for a
complex full likelihood.  The properties of the estimators found using
this scoring rule are compared with the full and pairwise maximum
likelihood estimators.  Two examples are discussed: the first a
stationary first order autoregressive model, and the second a first
order moving average model.  In the first example the pairwise method
produces good estimators; in contrast, in the second example this
method leads to poor estimators.  The opposite behaviour is observed
for the univariate and multivariate Hyv\"arinen estimators.  For the
moving average process, there can be a large gain in efficiency, as
compared to the pairwise likelihood method, by using the univariate or
multivariate Hyv\"arinen score.  For the autoregressive model, in
contrast, the Hyv\"arinen score methods suffer a loss of efficiency as
$|\phi|$ approaches $1$.  In both examples, a great improvement in the
performances of the minimum Hyv\"arinen score based on the Wishart
model is observed as the ratio $T/\nu$ becomes negligible.  It is
known that the algorithm used to generate a Wishart random matrix as
the sum-of-squares-and-products matrix of independent multivariate
normals is not efficient (see for example \citet[pag.150]{monte}).  A
question which arises is whether the inefficiency of this algorithm
might be affecting the observed behaviour of the multivariate
Hyv\"arinen score.  However, results not shown here reveal that no big
improvement arises if we generate directly from the Wishart
distribution, using for example the \texttt{rwish} function of the
\texttt{MCMCpack} package, which use the Bartlett's decomposition (see
\citet[p.~240]{Kollo}).  It is clear that the loss of efficiency
incurred in using the Hyv\"arinen scoring rule or pairwise likelihood
can be quite substantial, but this depends on the underlying model.
The multivariate Hyv\"arinen estimator has the apparent advantage over
the other estimators (apart from full maximum likelihood) of being
based on the sufficient statistic of the model; nevertheless the
univariate Hyv\"arinen methods shows good performance in terms both of
standard errors and efficiency.  The Hyv\"arinen scoring rule methods
may represent viable alternatives to the pairwise log-likelihood
approach for inference in high-dimensional models where the
computation of the normalizing constant is not feasible and the
pairwise likelihood leads to poor estimators.  In particular, the
multivariate Hyv\"arinen scoring rule may be convenient for studies in
which a large number of models with the same parameter should be
estimated.  It would be of interest to analyse the performance of the
univariate and the multivariate Hyv\"arinen scoring rule estimators
both when nuisance parameters are present and when interest focus on
the complete vector of parameters.
\section*{Acknowledgements}
This research was partially supported by a grant from the University
of Cagliari (Progetto di Ricerca Fondamentale o di Base 2012).  The
authors are grateful to Dr.\ Manuela Cattelan for helpful advice on
the numerical evaluation of the Godambe information.

\end{document}